\documentclass[preprint,floatfix,showpacs,preprintnumbers,amsmath,amssymb,showkeys]{revtex4}

\usepackage{graphicx}
\usepackage{dcolumn}
\usepackage{bm}
\newcommand{\mysection}[1]{\vspace{3mm} \par{\bf #1.}}

\begin{document}

\title{The strong coupling Kondo lattice model as a Fermi gas }
\author{Stellan \"Ostlund}  
\email{ostlund@physics.gu.se}
\affiliation{
G\"oteborg University \\ Gothenburg 41296, Sweden \\
}

\newcommand{\ellipsis}{...}
\newcommand{\cdop}[2]{c^{\dagger}_{#1}(#2) }           
\newcommand{\cop}[2]{c^{\phdagger}_{#1}( #2 )}
\newcommand{\ket}[1]{{ | \;  { #1 }  \; \rangle}}
\newcommand{\keta}[1]{{ | \;  { #1 }  \; \rangle}_{a}}
\newcommand{\ketc}[1]{{ | \;  { #1 }  \; \rangle}_{c}}
\newcommand{\chat}[2]{{\hat{c}}^{\dagger}_{#1}(#2)}           
\newcommand{\cP} {{\cal P}}
\newcommand{\dhat}[2]{{\hat{c}}^{\phdagger}_{#1}(#2)}           
\newcommand{\bra}[1]{{ \langle \; { #1 } \; | }}
\newcommand{\phdagger}{\phantom{\dagger}}
\newcommand{\expectation}[1]{\langle \;  #1 \; \rangle}
\newcommand{\e}[1]{\langle \;  #1 \; \rangle}
\newcommand{\nh}{\hat{n}}
\newcommand{\rp}{{r'}}
\newcommand{\half}{\frac{1}{2}}
\newcommand{\up}{\uparrow}
\newcommand{\down}{\downarrow}
\newcommand{\cdagger}[1]{c^{\dagger}_{#1} }           
\newcommand{\cS}{{\cal S}}
\date{\today}

\begin{abstract}
The strong coupling half-filled Kondo lattice model is an important
example of a strongly interacting dense Fermi system for which
conventional Fermi gas analysis has thus far failed.  We remedy this
by deriving an exact transformation that maps the model to a dilute gas
of weakly interacting electron and hole quasiparticles that can then be
analyzed by conventional dilute Fermi gas methods.  The quasiparticle
vacuum is a singlet Mott insulator for which the quasiparticle dynamics
are simple.  Since the transformation is exact,  the electron spectral
weight sum rules are obeyed exactly.  Subtleties in understanding the
behavior of electrons in the singlet Mott insulator can be reduced to
a fairly complicated but precise relation between quasiparticles and
bare electrons.  The theory of free quasiparticles can be interpreted
as an exactly solvable model for a singlet Mott insulator, providing
an exact model in which to explore the strong coupling regime of a singlet Kondo
insulator.

\end{abstract}
\maketitle

\mysection{Introduction}
A Fermi liquid is described by starting with an ideal non-interacting
Fermi gas. We then adiabatically ``switch on'' the interactions between particles. 
States of the interacting system are then identified with those 
of the non-interacting system.  Fermi quasiparticles have the same relation to Fermi liquid ground
state as the electron operators do to the vacuum; if $\ket{\Psi_G}\;$ is the Fermi liquid ground state, there should exist
a quasiparticle creation operator ${c^{\dagger}_{\beta}}\;$ so that $0 = {c_{\beta}}\,\ket{\Psi_G}\;$ ,  $ \{c_\alpha,  c^\dagger_\beta \} = \delta_{\alpha\beta} $ 
and 
$  H c^\dagger_{\alpha} \ket{\Psi_G}  = e_{\alpha} c^\dagger_{\alpha} \ket{\Psi_G} $.  
The spectral weight sum-rule follows directly from this.

Although the spectral weight sum rule is so fundamental it has been
surprisingly difficult to implement in approximate theories of the Kondo
lattice model.\cite{ref:eder} The reason is that the ground state is
a dense system of strongly interacting Fermions for which it has been
difficult to obtain results using traditional weak coupling methods. This
work demonstrates an exact transformation that can be utilized to
map the dense strongly interacting singlet Mott insulator to a weak
coupling dilute Fermi gas. This allows conventional
Fermi gas techniques to be applied to this class of models.

We begin by  considering a lattice of sites each
consisting of conduction band ``c'' electrons that can hop
between neighboring sites and core ``f'' electrons that are
confined.
At half filling  a total of two electrons tend to be located on the
site.  A potential favors single
occupancy of the ``f'' electron, forcing the second electron into the
conduction band.  The extended model consists of a lattice of such sites
with  only the ``c'' electrons hopping between neighbors.

There are four states  with one ``c'' and one ``f'' electron present.
The resulting triplet-singlet degeneracy is lifted by a spin exchange term
that gives the singlet lowest energy.\cite{ref:noproblem}
This behavior is most easily encoded in the Kondo lattice Hamiltonian
\cite{ref:oshikawa}
\begin{equation} H_{KLM} = \begin{array}[t]{l}  t \, \sum_{r\rp}  ( \cdop{c,\sigma}{r} \cop{c,\sigma}{\rp} + CC) \; + \\  \; 
 \sum_r  \, ( J \; S_c(r) \cdot S_f(r)  + U_f ( n_f(r) - 1 )^2  \, ) 
\end{array} .
\label{eq:kondohubbard} \end{equation}
The atomic ground state is given by 
$\ket{\Psi_G} = \frac{1}{\sqrt{2}}\,({c^{\dagger}_{c,\uparrow}}\,{c^{\dagger}_{f,\downarrow}} - {c^{\dagger}_{c,\downarrow}}\,{c^{\dagger}_{f,\uparrow}})\,\ket{0}.\;$ The full ground state is a product
of such two-electron singlets at each site  and is nondegenerate with
finite gap to the excited states.  The gap will persist for small values
of the hopping $ t $ and the local conduction electron will then simply
make virtual excursions to the neighboring sites. The system will remain
a singlet Mott insulator.

Before describing quasiparticles for nonzero $ t $ we begin with the
limit of zero hopping.  For the Hubbard impurity, there are four
charged spin half excitations on a single site, consisting of each of
the four states with three electrons.\cite{ref:oshikawa}
We first attempt to define a quasiparticle operator simply by
$  {c}^\dagger_{c,\sigma}  $ which one might hope adds an 
extra conduction electron to the singlet.

This ansatz immediately leads to problems. First we find ${c^{\dagger}_{c,\uparrow}}\,\ket{\Psi_G} = -\,\frac{1}{\sqrt{2}}\,{c^{\dagger}_{c,\uparrow}}{c^{\dagger}_{c,\downarrow}}{c^{\dagger}_{f,\uparrow}}\,\ket{0}.\;$ Although the operator indeed creates a state with an extra 
charge, this state is not normalized. 

A second problem is the following inequality:
${c^{\phdagger}_{c,\uparrow}}\,\ket{\Psi_G} = \frac{1}{\sqrt{2}}\,{c^{\dagger}_{f,\downarrow}}\,\ket{0}\ne 0\;$ which shows that the annihilation operator does
not annihilate the atomic ground state.  Finally, we note that  ${c^{\phdagger}_{c,\uparrow}}{c^{\dagger}_{c,\uparrow}}\,\ket{\Psi_G} = -\,\frac{1}{\sqrt{2}}\,{c^{\dagger}_{c,\downarrow}}\,{c^{\dagger}_{f,\uparrow}}\,\ket{0}\ne\ket{\Psi_G}\;$ so that this simple annihilation operator 
does not reconstruct the original singlet ground state.  We conclude that
none of the conditions we require of a proper quasiparticle operator
is fulfilled with this ansatz.  This simple analysis reveals the source of
the difficulty;  the ground state is ``entangled'' in the sense that
it cannot be created from the true vacuum by a simple product of bare
electron operators.

\mysection{Constructing correct local  quasiparticles} 
\label{sec:transformation}

We now discuss how to define proper quasiparticle operators.
The local Hilbert space consists of sixteen states which we 
label from $ 1 $  to $ 16 $.  First we list  the eight states with even
particle number ordered first by increasing particle number.
Degeneracies are split by total spin, and remaining degeneracies are
split by $ s_z$.  States with odd particle number 
are similarly assigned labels $ 9 $  to $ 16 $.

\begin{figure}[h]
\includegraphics[width=8cm]{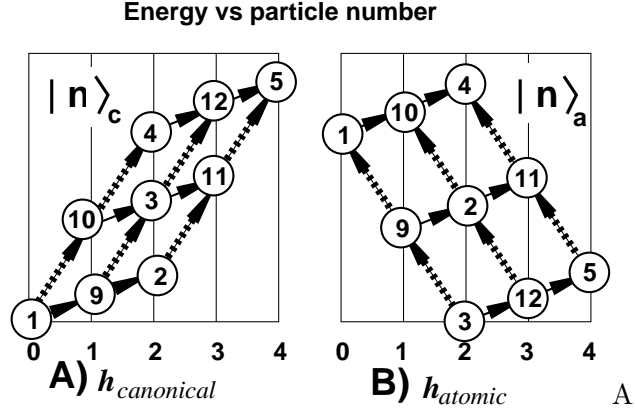}A
\caption{ In (A), the energy 
$  E =   n_tot - \frac{2}{3} ( n_f - n_c ) $ is plotted vs particle number
and labeled with the state that has the smallest value of $S^2 $ and $ S_z $.
A dotted arrow indicates a c-type creation operator is
used to create the state at the tip of the arrow from the state 
at the base. A a solid arrow indicates an f-type creation operator is
used. A solid-dotted combination indicates a singlet combination is used. 
In (B) a similar construction is used in an interacting model.
The dotted arrow here indicates a hole operator. 
}
\label{fig:fig1} 
\end{figure}

This ``canonical'' ordering is shown on an 
energy versus particle number diagram for the Hamiltonian $h_{canonical} = n_{tot} - \frac{2}{3}\,({n_{f}} - {n_{c}})\;$ in Figure \ref{fig:fig1}A.
On the horizontal axis is plotted 
$ n_{tot}  $.  On the vertical axis is the energy state labelled
with the state of minimum z-component of spin for each degenerate 
multiplet. In Figure \ref{fig:fig1}B is a hypothetical
diagram with an interacting Hamiltonian $ h_{atomic} $ 
favoring the Kondo singlet ground state.  To avoid cluttering the
figure, triplet states are not drawn.

 We denote $ \ket{n}_a \equiv \ket{n}_{atomic} $ states
in Fig. \ref{fig:fig1}B and 
$ \ket{n}_c \equiv \ket{n}_{canonical} $ states
states in Fig. \ref{fig:fig1}A.
   A precise description of these states is needed to continue.

The singlets are given by $\ketc{1} = 1\ket{0},\;$$\ketc{2} = {c^{\dagger}_{f,\uparrow}}\,{c^{\dagger}_{f,\downarrow}}\ket{0},\;$$\ketc{3} = \frac{1}{\sqrt{2}}\,({c^{\dagger}_{c,\uparrow}}\,{c^{\dagger}_{f,\downarrow}} - {c^{\dagger}_{c,\downarrow}}\,{c^{\dagger}_{f,\uparrow}})\ket{0},\;$$\ketc{4} = {c^{\dagger}_{c,\uparrow}}\,{c^{\dagger}_{c,\downarrow}}\ket{0},\;$$\ketc{5} = {c^{\dagger}_{c,\uparrow}}{c^{\dagger}_{c,\downarrow}}{c^{\dagger}_{f,\uparrow}}{c^{\dagger}_{f,\downarrow}}\ket{0}.\;$
The triplets numbered $ \ketc{6},\ketc{7} $ and $ \ketc{8} $ are
not shown in the diagram.

The four states of total spin $ -\half $  are given by: $\ketc{9} = {c^{\dagger}_{f,\downarrow}}\ket{0},\;$$\ketc{10} = {c^{\dagger}_{c,\downarrow}}\ket{0},\;$$\ketc{11} = {c^{\dagger}_{c,\downarrow}}{c^{\dagger}_{f,\uparrow}}{c^{\dagger}_{f,\downarrow}}\ket{0},\;$$\ketc{12} = {c^{\dagger}_{c,\uparrow}}{c^{\dagger}_{c,\downarrow}}{c^{\dagger}_{f,\downarrow}}\ket{0}.\;$ The states of total spin $ \half  $: are $\ketc{13} = {c^{\dagger}_{f,\uparrow}}\ket{0},\;$$\ketc{14} = {c^{\dagger}_{c,\uparrow}}\ket{0},\;$$\ketc{15} = {c^{\dagger}_{c,\uparrow}}{c^{\dagger}_{f,\uparrow}}{c^{\dagger}_{f,\downarrow}}\ket{0},\;$$\ketc{16} = {c^{\dagger}_{c,\uparrow}}{c^{\dagger}_{c,\downarrow}}{c^{\dagger}_{f,\uparrow}}\ket{0}.\;$ The  states with $ s_z = - \half $ are not labelled in
the figure. 

A local spin-symmetric Hamiltonian $H$
is block diagonal in the blocks enclosed by parentheses:
$ \left( [1],[2,3,4],[5]\right) ([6]),([7]),([8]) $
$ \left([9,10][11,12]\right) \left([13,14][15,16]\right) $.
If particle number is also conserved, $H$ 
is further block diagonal in the subblocks in square brackets. 


We now draw in Fig. \ref{fig:fig1}B a  similar diagram for a 
more complicated Kondo-Lattice like Hamiltonian with identical 
but permuted eigenvalues.  We assume it has the Kondo singlet ground 
state and lowest energy charge $ 2 \pm 1 $  states
consisting of zero or two c-electrons together with a single 
``f'' electron.   
 Eigenstates in Fig. \ref{fig:fig1}B  
are labelled according to the same ordering scheme.

Our goal is to construct a unitary transformation $ U $ that maps
the Kondo singlet ground state in Fig. \ref{fig:fig1}B to the
vacuum in Fig. \ref{fig:fig1}A and at the same time define
quasiparticle operators that create the states of charge $ 2 \, \pm \,  1 $
from the Kondo singlet.  The arrows in Fig.  \ref{fig:fig1}B
illustrate the action of the quasiparticle operators.  A dotted  hole
creation operator creates the singly charged states and or a solid
electron operator creates the triply charged states.  
We thus map the ground state $ \keta{3} $ to the 
vacuum $ \ketc{1} \equiv \ketc{G}  $ i.e. $ U \keta{3} =  \ketc{1} $.   The low energy state of one extra spin down electron 
$ \keta{12} $ must map to  $ \ketc{9} $; $ U \keta{12}  =  \ketc{9} $.   Similarly we demand $ U \keta{9}  =  \ketc{10} $ .   

Symmetries can now be used to completely specify $ U $.
We define  $ Q $ to be the combined particle-hole spin flip 
transformation.  This permutation is an inversion
of the picture in Fig. \ref{fig:fig1}A through the state $ \ketc{3} $.  
Let $ G_{ij} $ be the $ 16 \times 16 $ matrix that has all entries
zero except $ G_{ij} = 1 $.  It can be verified that  $Q = {G_{1,5}} + {G_{3,3}} + {G_{5,1}} + {G_{6,6}} + {G_{7,7}} + {G_{8,8}} + {G_{9,12}} + {G_{10,11}} + {G_{13,16}} + {G_{14,15}} - {G_{16,13}} - {G_{15,14}} - {G_{12,9}} - {G_{11,10}} - {G_{4,2}} - {G_{2,4}}.\;$  Signs are chosen to preserve spin.  

We next define $ K $ to be the transformation that exchanges the two
species of Fermions.  This corresponds to the permutation obtained by
exchanging the high and low energy state for each value of $ n_{tot}$ in
Fig. \ref{fig:fig1}A.  Minus signs are inserted to make the transformation 
an element of the group of continuous rotations.  We find $K = {G_{1,1}} + {G_{2,4}} + {G_{4,2}} + {G_{5,5}} + {G_{6,6}} + {G_{7,7}} + {G_{8,8}} + {G_{10,9}} + {G_{11,12}} + {G_{14,13}} + {G_{15,16}} - {G_{16,15}} - {G_{13,14}} - {G_{12,11}} - {G_{9,10}} - {G_{3,3}}.\;$ 

In order for $ U $ to preserve spin and also generate particle and 
hole operators from the original Fermions, we demand $ U $ obeys
the following:   $ [\,U,\,S\, ] = 0  $, $ \; n_{tot} \, U  =  U \, ( 2 + n_e - n_h ) $  
and $  U K = Q U $.  It can be verified that these constraints together with 
the demand that $ U $ be real and unitary results in \begin{equation}U = \begin{array}[t]{l}
{G_{1,4}} + {G_{2,3}} / \sqrt{2} + -\,{G_{2,5}} / \sqrt{2} + {G_{3,1}} \; \;  \text{+} \\
{G_{4,3}} / \sqrt{2} + {G_{4,5}} / \sqrt{2} + {G_{5,2}} + {G_{6,6}} \; \;  \text{+} \\
{G_{7,7}} + {G_{8,8}} + {G_{9,10}} + -\,{G_{10,12}} \; \;  \text{+} \\
{G_{11,11}} + {G_{12,9}} + {G_{13,14}} + -\,{G_{14,16}} \; \;  \text{+} \\
{G_{15,15}} + {G_{16,13}}
\end{array}\nonumber.\end{equation}
 This defines a new Hamiltonian  
$ h_{atomic}^\prime = U_r^\dagger h_{canonical} U_r $ which
has the energy diagram shown in Fig \ref{fig:fig1}B . 
The demand that $ U^\dagger \, n_{tot} \,  U = 2 + n_e - n_h $ shows
that $ U $ cannot be continued to the identity transformation which is
presumably relevant for $ J = 0 $. This observation adds support to the existence
of a quantum phase transition  for sufficiently large values of $ t/J$,
a subject which is outside the scope of this investigation.

From unitarity follows that 
$ \chat{e,\sigma}{r} \, \equiv \,  U_r^\dagger \cdop{c,\sigma}{r}U_r $  and
$ \chat{h,\sigma}{r} \, \equiv \,  U_r^\dagger \cdop{f,\sigma}{r}U_r $ preserve
the Fermi anticommutation relations. Inverting this formula and using 
the fact that  
$ U_r\keta{\Psi_G } \equiv U_r \keta{3}  =  \ketc{1}  \equiv \ketc{Vac} $ we find that
$ \cop{cf,\sigma}{r}\ketc{Vac} = 0 $ is equivalent 
to $  \dhat{eh,\sigma}{r} \keta{\Psi_G }  = 0 $. We have therefore obtained
a local unitary transformation that maps the original 
``c'' and ``f'' bare electron operators to quasiparticle operators 
that annihilate the Kondo singlet.  By construction 
the operators $ \chat{eh}{r} $ either add or remove exactly one
charge in each of the states and 
 $ \chat{e,\sigma}{r} $  is identified as
an electron quasiparticle operator 
and $ \chat{h,\sigma}{r} $  as a hole.

Using the method in Ref. \cite{ostlundhansson} and Ref. \cite{unpub:ostlund} we can 
explicitly write down the original operators 
$ c^\dagger_{cf} $ and $ c_{cf} $  as a polynomial of Fermion 
operators  $ \chat{eh,\sigma}{r}, \dhat{eh,\sigma}{r} $.  The result 
is 
\begin{equation}{c^{\dagger}_{c,\uparrow}} = \begin{array}[t]{l}
(\frac{1}{\sqrt{2}}\,{\hat{c}^{\phdagger}_{h,\downarrow}} - \frac{1}{\sqrt{2}}\,{\hat{c}^{\dagger}_{e,\uparrow}}) + \\
{\hat{c}^{\phdagger}_{h,\downarrow}}\,(\tau_1\,{n_{h,\uparrow}}\,{n_{e,\uparrow}} + \tau_2\,{n_{e,\downarrow}} + \tau_1\,{n_{h,\uparrow}}\,{n_{e,\downarrow}} + \tau_1\,{n_{e,\uparrow}}\,{n_{e,\downarrow}} + \frac{1}{2}\,{\hat{c}^{\dagger}_{h,\uparrow}}\,{\hat{c}^{\dagger}_{e,\downarrow}} - \tau_1\,{n_{e,\uparrow}} - \frac{1}{\sqrt{2}}\,{n_{h,\uparrow}}) \; \;  \text{+} \\
{\hat{c}^{\dagger}_{e,\uparrow}}\,(\tau_1\,{n_{h,\downarrow}} + \frac{-1}{2}\,{\hat{c}^{\phdagger}_{h,\uparrow}}\,{\hat{c}^{\phdagger}_{e,\downarrow}} + \frac{1}{\sqrt{2}}\,{n_{e,\downarrow}} - \tau_1\,{n_{h,\downarrow}}\,{n_{e,\downarrow}} - \tau_1\,{n_{h,\uparrow}}\,{n_{e,\downarrow}} - \tau_1\,{n_{h,\uparrow}}\,{n_{h,\downarrow}} - \tau_2\,{n_{h,\uparrow}}) \; \;  \text{+} \\
\end{array}\nonumber\end{equation}
where  $ \tau_{2}  =  ( 1 -   1/ \sqrt{2} ) $ and    $ \tau_{1} =  \half( \sqrt{2} - 1  )  $.  For the ``f'' electrons I find 
\begin{equation}{c^{\dagger}_{f,\uparrow}} = \begin{array}[t]{l}
{\hat{c}^{\dagger}_{e,\uparrow}}\,(\frac{1}{2}\,{n_{h,\downarrow}} + \frac{-1}{2}\,{n_{h,\uparrow}}\,{n_{h,\downarrow}} + {n_{e,\downarrow}} + \frac{-1}{2}\,{n_{h,\uparrow}}\,{n_{e,\downarrow}} + \frac{-1}{2}\,{n_{h,\downarrow}}\,{n_{e,\downarrow}} + \frac{-1}{2}\,{\hat{c}^{\phdagger}_{h,\uparrow}}\,{\hat{c}^{\phdagger}_{e,\downarrow}} + \frac{1}{\sqrt{2}}\,{\hat{c}^{\dagger}_{h,\uparrow}}\,{\hat{c}^{\dagger}_{e,\downarrow}}) \; \;  \text{+} \\
{\hat{c}^{\phdagger}_{h,\downarrow}}\,(\frac{-1}{2}\,{n_{e,\uparrow}} + \frac{1}{2}\,{n_{h,\uparrow}}\,{n_{e,\uparrow}} + \frac{1}{2}\,{n_{h,\uparrow}}\,{n_{e,\downarrow}} + \frac{1}{2}\,{n_{e,\uparrow}}\,{n_{e,\downarrow}} + \frac{1}{2}\,{\hat{c}^{\dagger}_{h,\uparrow}}\,{\hat{c}^{\dagger}_{e,\downarrow}} - \frac{1}{\sqrt{2}}\,{\hat{c}^{\phdagger}_{h,\uparrow}}\,{\hat{c}^{\phdagger}_{e,\downarrow}} - {n_{h,\uparrow}})
\end{array}\nonumber.\end{equation}
Similar formulas occur for the other spin component.
We note that  the ``f'' electron operators is constructed by third and higher
order electron and hole operators in contrast to the conduction electron
operators that have a linear coupling to quasiparticles.  Deriving or
even verifying these formulas is formidable without the help of a
computer.\cite{unpub:ostlund}

\mysection{Effective model for quasiparticles}
The transformation $ U $ is now used to define a global 
transformation by $ U_{global} = \prod_{r} U_{r} $.
Since $ U_{r} $ does not mix states of even and odd particle number we obtain 
$\{\cdop{i,\sigma}{r} , \cdop{j,\sigma}{\rp} \} = \delta_{i,j} \delta_{r,\rp} $ 
verifying that $ U $ is indeed a global canonical transformation.  

We now express the Kondo lattice model given by Eq. \ref{eq:kondohubbard}
in terms of electron and hole quasiparticle operators by  replacing
the  Fermi operators $  c_{cf,\sigma}(r) $ by their representation as
polynomials in $ \chat{eh,\sigma}{r} $.

By construction, the Kondo singlet is the quasiparticle vacuum in the
limit $ t \rightarrow  0 $.  For small values of $ t $, quasiparticles
can hop between adjacent sites. The resultant low energy charged charged
states that appear in the interacting ground state are therefore described
by  single quasiparticle  operators and the unitary transformation
guarantees these local states also have the correct energy.

For small $ t $ the states with local charge fluctuations  will have a
small amplitude in the interacting ground; the probability that a site
is occupied by more than one quasiparticle will be accordingly less.
The system should therefore be well approximated by a dilute fermi gas.

We  thus expand $ H_{KLM} $ in quasiparticle operators.  
Inserting the quasiparticle representation for the original fermions, normal 
ordering then truncating the Hamiltonian to second order we find\cite{ref:eder2}   
$ H_{KLM} = H_{KLM}^{free} + H_{KLM}^{interacting} $  
with
\begin{equation}\begin{array}[t]{l}
H_{KLM}^{free}  = \frac{3}{4}\,J\,(n_c + n_f - 1) \; \;  \text{+} \\
\frac{1}{2}\,t\Sigma_{s}\,(-1)^{\frac{1}{2} + s}\,(\chat{e,s}{r}\,\chat{h,-\,{s}}{\rp} - 
\dhat{e,s}{\rp}\,\dhat{h,-\,{s}}{r}) \; \;  \text{+} \\
\frac{1}{2}\,t\Sigma_{s}\,\chat{e,s}{r}\,\dhat{e,s}{\rp} -
 \chat{h,s}{r}\,\dhat{h,s}{\rp} \; \;  \text{+} \\
\frac{1}{2}\,t\Sigma_{s}\,\chat{e,s}{\rp}\,\dhat{e,s}{r} -
 \chat{h,s}{\rp}\,\dhat{h,s}{r} \; \;  \text{+} \\
-\,\frac{1}{2}\,t\Sigma_{s}\,(-1)^{\frac{1}{2} + s}\,(\dhat{h,s}{\rp}\,\dhat{e,-\,{s}}{r} -
 \chat{h,s}{r}\,\chat{e,-\,{s}}{\rp})
\end{array}\nonumber.\end{equation}
 
$ H_{KLM}^{free} $ can then be treated exactly by transforming
to momentum space and utilizing a conventional Bogoliubov transformation.\cite{unpub:ostlund}
With a chemical potential we compute the eigenvalues exactly.
We let $ e_k = \sum_{i} \cos( k \cdot \hat{i} )  $.  
The quasiparticle spectrum is then given by
 \label{eq:spec}  
$ E_k = \frac{3 J}{4} \Delta_k \pm  t e_k \pm \mu $ where 
$ \Delta_k =   \sqrt{ 1 + ( \frac{ 4 t e_k  }{ 3 J }  )^2  } $. 

It is reassuring that  $ U_f $ does not appear to quadratic
order; a $ J $ which is not small is what is crucial  in 
creating the Kondo singlet and low-energy states. 
For completeness, the entire onsite part of $ H_{KLM} $  is given by  
$H^{interacting}_{KLM} =  $ 
\begin{equation}\begin{array}[t]{l}
(n_e (n_e - 1) n_h + n_e n_h ( n_h - 1 ))(\frac{3}{16}\,J + \frac{-1}{4}\,U) \; \;  \text{+} \\
(n_e ( n_e - 1 ) + n_h ( n_h - 1 ))(\frac{-3}{8}\,J + \frac{1}{2}\,U) \; \;  \text{+} \\
((S_e \cdot S_h )\,(\frac{1}{4}\,J - U) + n_e n_h\,(\frac{-9}{16}\,J + \frac{1}{4}\,U))
\end{array}\nonumber.\end{equation}
High order terms in the hopping become too lengthy to write
out here, but are unimportant for the  small $ t $ physics
since not only is the prefactor $ t $ small, but also,
as we shall see , the fourth order Fermion operators 
become unimportant since the Fermi gas is dilute.

\begin{figure}[h]
\includegraphics[trim = 5mm 30mm 0mm 10mm , clip, width=6cm]{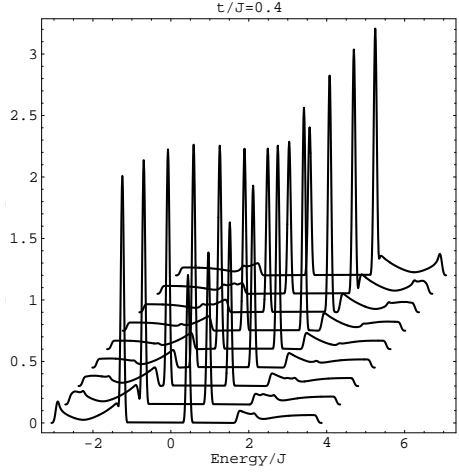}
\caption{ Square root of the spectral weight for c-electrons $ A_c(k,\omega) $ 
is plotted for $ H_{KLM}^{free} $ for $t/J = 0.4\;$ and $ t = 1 $ in one dimension.
To enable plotting the coherent and incoherent parts together,
$ A_c(k,\omega) $ has been smoothed by a Gaussian in $ \omega $ before taking 
the square root.  Plots are shown for $ k = 0 $ to $ k = \pi $.  }
\label{fig:fig2}
\end{figure}

\mysection{Spectral weights}
Since the local electron operator is represented in terms of local 
quasiparticles the spectral weights can be computed for $ H^{free}_{KLM}$.
As an example, the results are plotted in in Fig. \ref{fig:fig2} where
the spectral weight for the c-electron
$ A_c(k,\omega) $ is plotted as a function  of $ \omega  $
several values of $ k $ from $ 0 $ to $ \pi $. The quasiparticle
coherent peak is seen as a broadened $ \delta $ function. The
incoherent structure is comprised of three-quasiparticle
contributions. Not shown are the five-quasiparticle contributions that 
represent less than 1 \% of the total spectral weight.
We observe that $ A_c(k + \pi ,\omega) =  A_c( k , -\omega ) $
a consequence of the particle-hole symmetry of  
$ H_{KLM} $.

\mysection{Numerical results}
The quality of the wave function as an approximation to the Kondo
Lattice ground state can be measured by the overlap of the ground state
with the Hilbert space of $ H_{KLM} $.    Let
$  \alpha  = \half   \int \! \! \! \int \! \! \! \int_{T^d}  ( 1 - \Delta_k^{-1}
  ) \; (  \frac{d \, \theta}{2 \pi } )^d
\approx  \frac{2 d }{9} \lambda^2  + O( \lambda^4) $ where $ \lambda = t/J $
and $ d $ is the dimension.  The deviation from unity is measured by $ \delta_P = \e{(1-n_f)^2} $
which given exactly by  
$ \delta_P = \alpha^2 ( 3 - 2 \alpha ) $ demonstrating
convergence like $ (t/J)^4 $ for small $ t$ .  We also find
$ \expectation{n_e + n_h} = 4  \; \alpha $ which
verifies that the number of quasiparticles remains
surprisingly small even as the Kondo lattice 
model approaches a phase transition.

A more challenging quantity to compute is the expectation 
value of the complete full Kondo Lattice Hamiltonian  with
$ U_f = 0 $ 
in the ground state of $ H_{KLM}^{free} $. This quantity
is important since it is a rigorous upper bound to the {\it true } Kondo 
lattice  ground state energy.  We define 
$ \beta =   \; \int \! \! \! \int \! \! \! \int_{T^d}   e_k^{2} / \Delta_k
  \; (  \frac{d \, \theta}{2 \pi } )^d  
\approx  \frac{d}{2} - \frac{d}{3} \lambda^2  + O( t^4 ) $
and  find\cite{unpub:ostlund} 
\begin{equation}\begin{array}[t]{l}
\expectation{H_{KLM} }_{free} = 
\frac{-3}{4} + 3\,\alpha + \frac{-15}{4}\,\alpha^{2} + \frac{3}{2}\,\alpha^{3} +
 \frac{-8}{3}\,\beta\,\lambda^{2} \; \;  \text{+} \\
\alpha^{3}\,\beta\,\lambda^{2}\,(72 + -48\,\sqrt{2}) +
 \alpha\,\beta^{3}\,d^{-2}\,\lambda^{4}\,(\frac{64}{3} +
 \frac{-128}{9}\,\sqrt{2}) \; \;  \text{+} \\
\alpha\,\beta\,\lambda^{2}\,(16 + -8\,\sqrt{2}) +
 \beta^{5}\,d^{-4}\,\lambda^{6}\,(\frac{-64}{27} +
 \frac{128}{81}\,\sqrt{2}) \; \;  \text{+} \\
\beta^{3}\,d^{-2}\,\lambda^{4}\,(\frac{-64}{9} +
 \frac{32}{9}\,\sqrt{2}) + \alpha^{2}\,\beta^{3}\,d^{-2}\,\lambda^{4}\,(\frac{-64}{3} +
 \frac{128}{9}\,\sqrt{2}) \; \;  \text{+} \\
\alpha^{4}\,\beta\,\lambda^{2}\,(-36 + 24\,\sqrt{2}) +
 \alpha^{2}\,\beta\,\lambda^{2}\,(-52 + 32\,\sqrt{2})
\end{array}\nonumber
.
\end{equation}
 
This can be compared to high order series expansions.\cite{oitmaa}
We find our result is at most about 3 \% above the
series result, converging very rapidly for small values of $ \lambda $.
Errors are bounded by $ .61 \, \lambda^4 $ for $ d = 1 $,
$ 2.25  \,\lambda^4 $ for $ d = 2 $ and $ 4.9 \, \lambda^4 $ for 
$ d = 3 $ in the entire domain of convergence of the power series.  

\mysection{Conclusions}
An exact canonical transformation is derived that maps bare electron
operators to quasiparticle electron and hole operators.  This allows  us
to approximate the Kondo lattice model as a dilute Fermi gas for large
values of the Kondo coupling.  Properties of the Fermi gas can then
be computed exactly.  The overlap per site between the free Fermi gas
ground state and the Hilbert space of the Kondo Lattice model is 
$ ( 1 - O( t/J)^4 ) $ and ground state energies are also accurate to this order.
Spectral weights can be computed that exactly obey the sum rules.

Together with the exact relation between quasiparticle operators, the
free quasiparticle  Hamiltonian represents an exactly solvable
model of a singlet Mott insulator. The difficulty of understanding
electrons in this Mott insulator is encoded in the complicated relation
between electrons and quasiparticles rather than any subtlety in the
quasiparticle dynamics.


\begin{thebibliography}{8}
\bibitem{ref:noproblem}
We could have introduced a hopping instead of a spin exchange, leading
to an effective spin exchange $ J_{eff} = V^2/U $; Introducing more
parameters makes the subsequent analysis more complicated but it in
principle it can still be carried out with the same rigor.
\bibitem{ref:eder2}
This Hamiltonian is formally identical to that found in
Ref. \cite{ref:eder} but in that work, the electron and hole
operators represent constrained fermions yielding a number
of difficulties that are circumvented in the present work.
\end{thebibliography}
\end{document}